\documentclass{article}
\usepackage{spconf,amsmath,graphicx,hyperref}
\usepackage{enumitem}
\usepackage{algorithm}
\usepackage{algpseudocode}
\usepackage{amsmath,amssymb,amsfonts}
\usepackage{booktabs}
\usepackage{amsmath}
\usepackage[utf8]{inputenc}
\usepackage{color,xcolor}
\usepackage{colortbl}

\usepackage{etoolbox}

\setlist[itemize]{leftmargin=*, topsep=2pt, itemsep=2pt, parsep=0pt, partopsep=0pt}
\title{Co-initialization of control filter and secondary path via meta-learning for active noise control}
\name{Ziyi Yang$^{1}$, Li Rao$^{2}$, Zhengding Luo$^{1}$\textsuperscript{*}, Dongyuan Shi$^{3}$, Qirui Huang$^{1}$, and Woon-Seng Gan$^{1}$
\thanks{*Corresponding author: Zhengding Luo.
The code will be available at \href{https://github.com/yzyzieee/ICASSP26_co-init_meta-learning}{https://github.com/yzyzieee/ICASSP26\_co-init\_meta-learning}}}
\address{$^{1}$Smart Nation TRANS Lab, School of Electrical and Electronic Engineering,\\ Nanyang Technological University, Singapore\\
         $^{2}$Key Laboratory of Modern Acoustics, Institute of Acoustics, Nanjing University, Nanjing, China\\
         $^{3}$Northwestern Polytechnical University, Xi'an, China\\}
\begin{document}
%
\maketitle
\begin{abstract}
Active noise control (ANC) must adapt quickly when the acoustic environment changes, yet early performance is largely dictated by initialization. We address this with a Model-Agnostic Meta-Learning (MAML) co-initialization that jointly sets the control filter and the secondary-path model for FxLMS-based ANC while keeping the runtime algorithm unchanged. The initializer is pre-trained on a small set of measured paths using short two-phase inner loops that mimic identification followed by residual-noise reduction, and is applied by simply setting the learned initial coefficients. In an online secondary path modeling FxLMS testbed, it yields lower early-stage error, shorter time-to-target, reduced auxiliary-noise energy, and faster recovery after path changes than a baseline without re-initialization. The method provides a simple fast start for feedforward ANC under environment changes, requiring a small set of paths to pre-train.
\end{abstract}


\vspace*{-0.2cm}
\section{Introduction}
\vspace*{-0.3cm}
\label{sec:intro}

Active noise control (ANC) is widely used to reduce low-frequency noise by driving a secondary source to cancel the disturbance \cite{nelson1994active,sm1996active}. While recent data-driven and machine-learning approaches have been explored for ANC \cite{luo2023delayless,wang2025transferable,zhang2021deep}, the filtered-x LMS (FxLMS) framework remains the standard method behind many practical ANC systems across open windows, wearables, and openspace \cite{zhang2018wave,lam2020active,yang2023active,kuo2006active}. In FxLMS, the filtered reference is generated using an estimated secondary-path response. If it departs from the true plant, convergence slows and stability margins shrink \cite{kajikawa2012recent}. At the same time, a controller that starts from zero produces little effective cancellation until sufficient gradient information accumulates \cite{elliott2000signal}. As a result, start-up behavior depends on both the secondary-path assumption and the control-filter initialization.

When the acoustic environment changes, a common approach is adding online secondary-path modeling (OSPM), which injects auxiliary noise to identify the secondary path during ANC operation \cite{zhang2005comparison}. In practice, only low power of auxiliary noise can be used, so the identification SNR is low and the first gradient steps are noisy. As a result, start-up is slow and recovery after path changes is sluggish. Power-scheduling variants help but do not remove this limitation. \cite{zhang2003robust,hu2019online}.

Fast-recovery strategies for abrupt system changes have been explored in adaptive filtering, echo cancellation, and ANC\cite{lopes2017close,yang2017frequency,monteiro2017accelerating}. However, in ANC these efforts primarily focus on update mechanisms rather than on initialization of the control filter. Typical deployments start from zeros or nominal values, or carry over previous coefficients, while most research effort optimizes the FxLMS/OSPM mechanism (step-size or structure) rather than the starting point that governs the first moments after a change \cite{kajikawa2012recent,sun2020realistic,schumacher2011active}. Recently, meta-learning has emerged as a way to learn initializations that adapt in a few gradient steps from little data \cite{finn2017model,hospedales2021meta}, and first reports in ANC show faster early convergence when only the control filter is initialized \cite{shi2021fast}.

This paper addresses the problem of initialization for feedforward ANC under acoustic environment changes. We introduce a co-initialization for FxLMS-based algorithms that jointly initializes the control filter and the secondary-path model based on Model-Agnostic Meta-Learning (MAML). The initializer is pre-trained on a small set of measured paths with short two-phase inner loops (identification followed by residual-noise reduction) and is deployed by setting the learned initial coefficients while keeping the adaptive updates unchanged. Under an OSPM–FxLMS \emph{evaluation setup} with measured in-ear headphone paths~\cite{liebich2019acoustic}, the pre-trained initializer adapts faster to environment changes: it reaches the target error sooner, maintains lower early-stage residuals, requires less auxiliary-noise energy, and recovers more quickly after environment switches than a baseline without reinitialization but with identical updates. A data study further reveals that the diversity of training paths enhances the initializer, with secondary-path diversity exerting the stronger effect. The main contributions of this work are:
\begin{itemize}[label=\textbullet]
 \item \textbf{Few-shot meta-learned co-initialization for FxLMS-based ANC}: from a small set of measured paths, we jointly learn the initial controller and secondary-path model; at deployment, we simply set these learned initial values and keep the runtime adaptive updates unchanged.
  \item \textbf{Fast start and robust adaptation under environment changes}: the initializer reaches the target error sooner, maintains lower early residuals, reduces auxiliary-noise energy, and recovers faster after path switches, with generalization to similar unseen acoustic conditions.
\end{itemize}


\vspace*{-0.3cm}
\section{Online Secondary Path Modeling (OSPM)}
\label{subsec:zhang}
\begin{figure}[t]
  \centering
  \includegraphics[width=\linewidth]{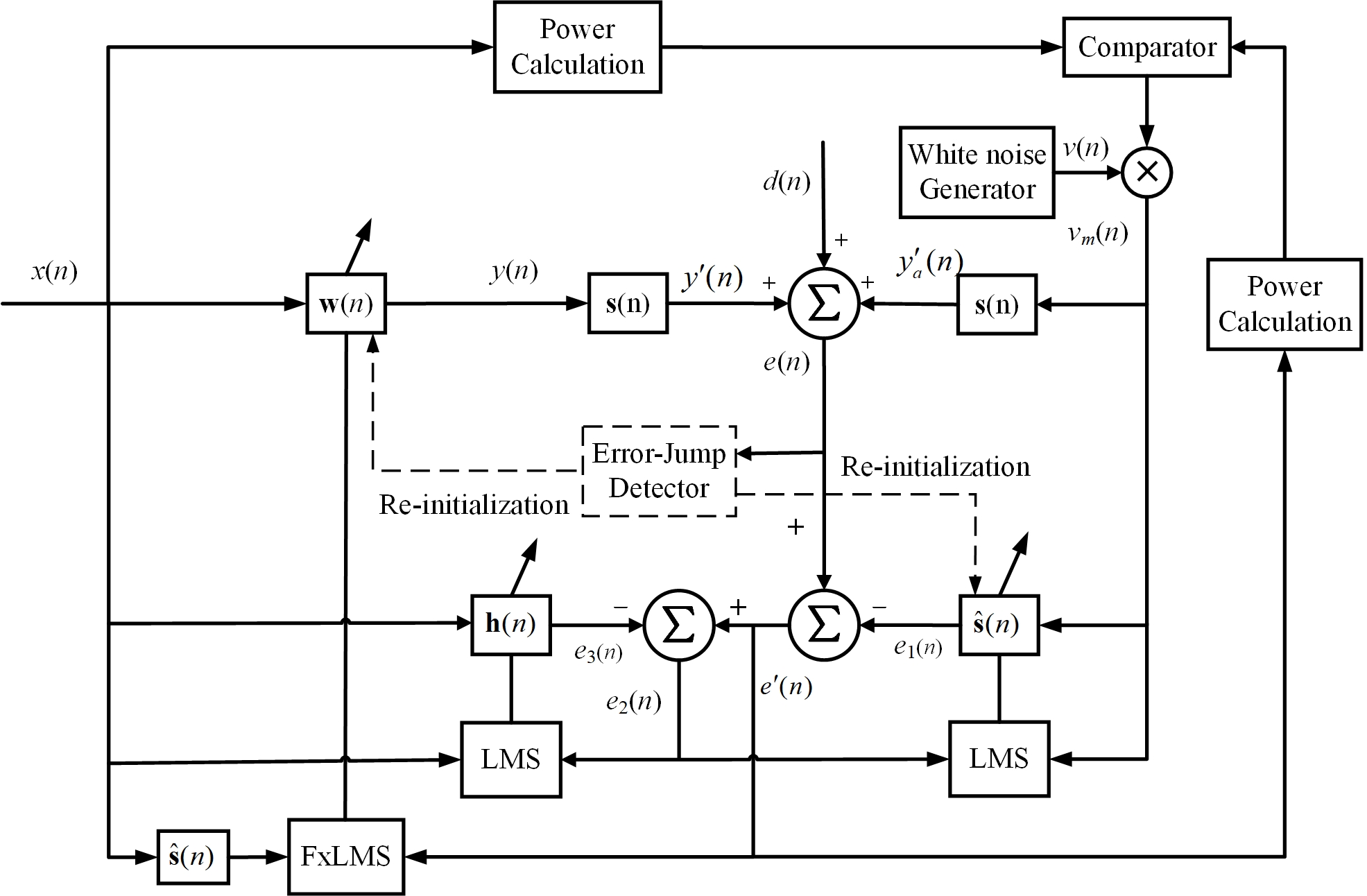}\vspace*{-0.6cm}
  \caption{Modified cross–updated online secondary–path modeling with auxiliary noise (with error jump detector). Dashed arrows indicate re–initialization and scheduling actions.}
  \label{fig:zhang}
\end{figure}

\vspace*{-0.3cm}
We adopt the cross-updated OSPM scheme of \cite{zhang2005comparison} and add a lightweight error-jump detector for re-initialization. The overall signal flow are shown in \autoref{fig:zhang}.

Let $x(n)$ be the reference, $d(n)$ the disturbance, and $v(n)$ a zero-mean white auxiliary noise. Its scaled output is $v_m(n)$. All FIRs are causal column vectors. With
$\mathbf{x}_w(n)=[x(n),\ldots,x(n-L_w+1)]^\top$,
$\mathbf{x}_s(n)=[x(n),\ldots,x(n-L_s+1)]^\top$, and
$\mathbf{u}(n)=[v_m(n),\ldots,v_m(n-L_s+1)]^\top$,
denote the true secondary path $\mathbf{s}\!\in\!\mathbb{R}^{L_s}$, its estimate $\hat{\mathbf{s}}$, the control filter $\mathbf{w}\!\in\!\mathbb{R}^{L_w}$, and the auxiliary canceller $\mathbf{h}\!\in\!\mathbb{R}^{L_s}$ (driven by a delayed stack $\mathbf{x}_h(n)$). The control output is $u(n)=\mathbf{w}^\top\mathbf{x}_w(n)$ and its last $L_s$ samples form $\mathbf{u}_w(n)$. The microphone residual and the auxiliary-cleaned error are
\begin{equation}
\setlength{\abovedisplayskip}{2pt}
\setlength{\belowdisplayskip}{2pt}
e(n)=d(n)+\mathbf{s}^\top\mathbf{u}_w(n)+\mathbf{s}^\top\mathbf{u}(n), \label{eq:e-model-A}
\end{equation}
and the cleaned error after removing the modeled auxiliary component is
\vspace*{-0.4cm}
\begin{equation}
e'(n)=e(n)-\hat{\mathbf{s}}^\top\mathbf{u}(n). \label{eq:e-clean-A}
\end{equation}
\par\vspace*{-0.25cm}\noindent
The three adaptive filters are updated in one pass as
\vspace*{-0.15cm}
\begin{align}
\hat{\mathbf{s}}(n{+}1)&=\hat{\mathbf{s}}(n)+\mu_s\,\mathbf{u}(n)\big[e'(n)-\mathbf{h}^\top(n)\mathbf{x}_h(n)\big], \label{eq:s-update-A}\\
\mathbf{h}(n{+}1)&=\mathbf{h}(n)+\mu_h\,\mathbf{x}_h(n)\big[e'(n)-\mathbf{h}^\top(n)\mathbf{x}_h(n)\big], \label{eq:h-update-A}\\
\mathbf{w}(n{+}1)&=\mathbf{w}(n)+\mu_w\,\tilde{\mathbf{x}}(n)\,e'(n), \label{eq:w-update-A}
\end{align}
\par\vspace*{-0.2cm}\noindent
where $\tilde{\mathbf{x}}(n)=[\,\tilde{x}(n),\,\ldots,\,\tilde{x}(n{-}L_w{+}1)\,]^\top$ and $\tilde{x}(n)=\hat{\mathbf{s}}^\top\mathbf{x}_s(n)$.
which ensures the control filter is always driven by the auxiliary–cancelled error while the secondary–path estimator uses that same error further cleaned by $\mathbf{h}$. To scale the auxiliary noise level, we adopt the  amplitude schedule\cite{zhang2003robust} with $0{<}\alpha{<}1$: 
\vspace*{-0.2cm}
\begin{equation}
P_x(n)=\alpha P_x(n{-}1)+(1{-}\alpha)\,x^2(n),
\label{eq:ema-power-x}
\end{equation}
\begin{equation}
P_{e'}(n)=\alpha P_{e'}(n{-}1)+(1{-}\alpha)\,e'^2(n),
\label{eq:ema-power-e}
\end{equation}
the instantaneous auxiliary signal is
\begin{equation}
v_m(n)=c_{\rm aux}\,v(n)\,\sqrt{\min\{P_x(n),\,P_{e'}(n)\}+\varepsilon},
\label{eq:vm-schedule-final}
\end{equation}
\vspace*{-0.0cm}
where $v(n)\!\sim\!\mathcal{N}(0,1)$ and $\varepsilon\!>\!0$ prevents vanishing injection. \\
Finally, we attach an error–jump detector that refreshes the initialization instantaneously upon a suspected acoustic environment change. We monitor the canceller norm $J_h(n)=\|\mathbf{h}(n)\|_2$ with a short look-back $M$ and threshold $\gamma_h$; if $\Delta J_h(n)=J_h(n)-J_h(n\!-\!M)>\gamma_h$, we instantly reset $\mathbf{h}\!\leftarrow\!\mathbf{0}$ and $(\hat{\mathbf{s}},\mathbf{w})\!\leftarrow\!(\boldsymbol{\Psi},\boldsymbol{\Phi})$, which are the meta-initializations defined in the \autoref{subsec:maml}. After that, the LMS/FxLMS updates \eqref{eq:s-update-A}–\eqref{eq:w-update-A} resume unchanged.

\vspace*{-0.4cm}
\section{Co-initialization for Feedforward ANC based on Meta Learning}
\label{subsec:maml}
\vspace*{-0.3cm}

In the ANC situation, each acoustic condition is treated as a task with discrete-time reference, disturbance, and the task secondary-path impulse response $\mathbf{s}\!\in\!\mathbb{R}^{L_s}$ (available only during meta-training).As summarized by the pseudocode in Algorithm \autoref{alg:maml_anc_sp}, 2 meta-parameters are learned: the control filter initialization $\boldsymbol{\Phi}\!\in\!\mathbb{R}^{L_w}$ and the secondary-path initialization $\boldsymbol{\Psi}\!\in\!\mathbb{R}^{L_s}$.The inner loop comprises Phase A (secondary-path update) and Phase B (control filter update), followed by validation on a separate segment and an across-task meta update. Inner-loop adaptation runs on short segments drawn from the input tracks $\mathbf{x}_c,\mathbf{d}_c$,  followed by validation on a separate segment and an across-task meta update. Scalars $\mu_w,\mu_s$ denote inner-loop step sizes, $\alpha_w,\alpha_s$ denote meta learning rates, and $\lambda_w,\lambda_s\!\in\!(0,1]$ are forgetting factors for validation losses.

\vspace*{-0.3cm}
\subsection{Inner Loop — Training (per sampled task)}
\vspace*{-0.3cm}
A contiguous segment $\{x(n),d(n)\}$ is sampled. Initialization sets
$\mathbf{w}\!\leftarrow\!\boldsymbol{\Phi}$ and $\hat{\mathbf{s}}\!\leftarrow\!\boldsymbol{\Psi}$.

1) Phase A: Secondary path update ($T_A$ steps).
Let $L_s$ denote the secondary-path filter length and let $v(n)$ be the auxiliary-noise used to excite the secondary path during training. Define the identification output as $y_{\mathrm{id}}(n)=\mathbf{s}^{\top}\mathbf{u}(n)$ with $\mathbf{u}(n)=[\,v(n),\,v(n\!-\!1),\,\ldots,\,v(n\!-\!L_s\!+\!1)\,]^{\!\top}$.  
At each step, compute the error
\vspace*{-0.4cm}
\begin{equation}
e_s(n)=y_{\mathrm{id}}(n)-\hat{\mathbf{s}}^{\top}\mathbf{u}(n),
\label{eq:yid_es}
\end{equation}
\par\vspace*{-0.3cm}\noindent
and apply the LMS recursion
\vspace*{-0.15cm}
\begin{equation}
\hat{\mathbf{s}}^{(t+1)}=\hat{\mathbf{s}}^{(t)}+\mu_s\,e_s(t)\,\mathbf{u}(t),
\qquad t=0,\ldots,T_A-1 .
\label{eq:s_update}
\end{equation}
\vspace*{-0.5cm}

2) Phase B: Control filter update ($T_B$ steps).
Define the filtered reference inline as $\tilde{x}(n)=\hat{\mathbf{s}}^{\top}\mathbf{x}_s(n)$ with $\mathbf{x}_s(n)=[\,x(n),\,x(n\!-\!1),\,\ldots,\,x(n\!-\!L_s\!+\!1)\,]^{\!\top}$, and let the $L_w$-tap control filter regressor be $\mathbf{x}'(m)=[\,\tilde{x}(m),\,\tilde{x}(m\!-\!1),\,\ldots,\,\tilde{x}(m\!-\!L_w\!+\!1)\,]^{\!\top}$.  
The residual and the FxLMS recursion are
\vspace*{-0.2cm}
\begin{equation}
e(m)=d(m)-\mathbf{w}^{\top}\mathbf{x}'(m),
\label{eq:w_err}
\end{equation}
\begin{equation}
\mathbf{w}^{(t+1)}=\mathbf{w}^{(t)}+\mu_w\,e(t)\,\mathbf{x}'(t),
\qquad t=0,\ldots,T_B-1 .
\label{eq:w_update}
\end{equation}
\vspace*{-0.1cm}
Phase~A and Phase~B use disjoint time indices within the inner segment.

\vspace*{-0.2cm}
\subsection{Inner Loop — Testing}
\vspace*{-0.1cm}
A new validation batch $\{x^{\dagger}(k),d^{\dagger}(k)\}$ from the same task is drawn, while $\hat{\mathbf{s}}$ and $\mathbf{w}$ are frozen. The validation errors are
\vspace*{-0.1cm}
\begin{equation}
e_s^{\dagger}(k) \;=\; \mathbf{s}^{\top}\mathbf{u}^{\dagger}(k) - \hat{\mathbf{s}}^{\top}\mathbf{u}^{\dagger}(k),
\qquad k=n-t_s ,
\label{eq:sp_meta_err_dag}
\end{equation}
\begin{equation}
e^{\dagger}(k) \;=\; d^{\dagger}(k) - \mathbf{w}^{\top}\mathbf{x}'^{\dagger}(k),
\qquad k=n-t_w .
\label{eq:w_meta_err_dag}
\end{equation}
\vspace*{-0.1cm}
The meta-gradients are vector accumulations ( $\Delta\boldsymbol{\Psi}\!\in\!\mathbb{R}^{L_s}$, $\Delta\boldsymbol{\Phi}\!\in\!\mathbb{R}^{L_w}$ ) over the most-recent validation samples with forgetting factors $\lambda_s,\lambda_w\!\in\!(0,1]$:
\vspace*{-0.4cm}
\begin{equation}
\Delta\boldsymbol{\Psi}
\;=\; \sum_{t_s=0}^{N_s-1} \lambda_s^{\,t_s}\,
e_s^{\dagger}(n-t_s)\,\mathbf{u}^{\dagger}(n-t_s) ,
\label{eq:psi_sum}
\end{equation}
\begin{equation}
\Delta\boldsymbol{\Phi}
\;=\; \sum_{t_w=0}^{N_w-1} \lambda_w^{\,t_w}\,
e^{\dagger}(n-t_w)\,\mathbf{x}'^{\dagger}(n-t_w) .
\label{eq:phi_sum}
\end{equation}

\vspace*{-0.4cm}
\subsection{Across-task MAML update}
\vspace*{-0.2cm}
Given the task–level meta-gradients \eqref{eq:psi_sum}–\eqref{eq:phi_sum}, the meta-parameters are updated by a simultaneous gradient step
\vspace*{-0.2cm}
\begin{equation}
\label{eq:maml_update}
\boldsymbol{\Phi}^{(i+1)} \;=\; \boldsymbol{\Phi}^{(i)} + \alpha_w\,\Delta\boldsymbol{\Phi},
\qquad
\boldsymbol{\Psi}^{(i+1)} \;=\; \boldsymbol{\Psi}^{(i)} + \alpha_s\,\Delta\boldsymbol{\Psi},
\end{equation}
with learning rates $\alpha_w,\alpha_s>0$. After the update, the accumulators are reinitialized to zero, $\Delta\boldsymbol{\Phi}\!\leftarrow\!\mathbf{0}$ and $\Delta\boldsymbol{\Psi}\!\leftarrow\!\mathbf{0}$, before drawing the next task. Iterating this procedure for $i=0,\ldots,K-1$ across tasks yields a co-initialization $(\boldsymbol{\Phi},\boldsymbol{\Psi})$ that supports few-step secondary-path identification and residual-noise reduction under new conditions.

\begin{algorithm}[t]
\caption{Co–initialization of control filter and secondary path (SP) for ANC via MAML}
\label{alg:maml_anc_sp}
\begin{algorithmic}[1]
\State \textbf{Input:} reference track $\mathbf{x}_c$, disturbance track $\mathbf{d}_c$, task secondary path $\mathbf{s}$ (for training)
\State \textbf{Initialize:} meta control filter $\boldsymbol{\Phi}$, meta SP model $\boldsymbol{\Psi}$; inner steps $T_A,T_B$, test length $N$; step sizes $\mu_w,\mu_s$; forgetting factor $\lambda$; meta rates $\alpha_w,\alpha_s$; total epochs $K$
\vspace{2pt}
\For{epoch $i=1$ to $K$} \Comment{across tasks}
\State \textbf{/* Inner Loop - Training */}
  \State \textbf{Sample} a segment $\{x(n),d(n)\}$ from $(\mathbf{x}_c,\mathbf{d}_c)$
  \State $\,\mathbf{w}\leftarrow\boldsymbol{\Phi}$,\; $\hat{\mathbf{s}}\leftarrow\boldsymbol{\Psi}$ \Comment{inner initialization}
    \vspace{2pt}
    \State \textbf{/* Phase A: Secondary Path Update */}
      \State $\mathbf{u}(n)$ is samples of the reference to excite the SP,
      \State $y_{\!id}(n)=\mathbf{s}^\top\mathbf{u}(n)$,\quad $e_s(t)=y_{\!id}(n)-\hat{\mathbf{s}}^\top\mathbf{u}(n)$
      \State $\hat{\mathbf{s}}\leftarrow\hat{\mathbf{s}}+\mu_s\,e_s(n)\,\mathbf{u}(n)$
    \vspace{2pt}
    \State \textbf{/* Phase B: Control filter Update */}
      \State $\mathbf{x}'(m)$ is samples of the filtered reference from $\hat{\mathbf{s}}$,
      \State $e(m)=d(m)-\mathbf{w}^\top\mathbf{x}'(m)$
      \State $\mathbf{w}\leftarrow\mathbf{w}+\mu_w\,e(m)\,\mathbf{x}'(m)$
  \vspace{2pt}
  \State \textbf{/* Inner Loop - Testing */}
  \State $\Delta\boldsymbol{\Phi}\leftarrow\mathbf{0}$,\; $\Delta\boldsymbol{\Psi}\leftarrow\mathbf{0}$
  \For{$k_s=0$ to $N_s-1$}  
  \State $e_s^{\dagger}(k_s) \gets \mathbf{s}^\top\mathbf{u}^{\dagger}(k_s)-\hat{\mathbf{s}}^\top\mathbf{u}^{\dagger}(k_s)$
  \State $\Delta\boldsymbol{\Psi}\,{+}{=}\,\lambda_s^{\,k_s}\,e_s^{\dagger}(k_s)\,\mathbf{u}^{\dagger}(k_s)$
\EndFor

\For{$k_w=0$ to $N_w-1$}  
  \State $e^{\dagger}(k_w) \gets d^{\dagger}(k_w)-\mathbf{w}^\top\mathbf{x}'^{\dagger}(k_w)$
  \State $\Delta\boldsymbol{\Phi}\,{+}{=}\,\lambda_w^{\,k_w}\,e^{\dagger}(k_w)\,\mathbf{x}'^{\dagger}(k_w)$
\EndFor
  \State \textbf{/* MAML update */}
  \State $\boldsymbol{\Phi}\leftarrow\boldsymbol{\Phi}+\alpha_w\,\Delta\boldsymbol{\Phi}$,\quad
         $\boldsymbol{\Psi}\leftarrow\boldsymbol{\Psi}+\alpha_s\,\Delta\boldsymbol{\Psi}$
\EndFor
\end{algorithmic}
\end{algorithm}


\vspace*{-0.3cm}
\section{Simulation}
\label{sec:data}
\vspace*{-0.3cm}
\subsection{Path Dataset and Diversity Evaluation}
\label{pathdataset}
\vspace*{-0.25cm}
All simulations use measured in-ear primary and secondary paths from the RWTH Aachen IKS \textsc{PANDAR} database \cite{liebich2019acoustic}. In total, 46 primary–secondary path pairs are available.Measurements used Bose QC20 in-ear headphones, and subject-specific responses were captured in an acoustic booth. The set covers 23 listeners with 3 wearing conditions (normal, slightly loose, loose). Signals are resampled to $F_s{=}16$\,kHz for runtime, and the noise is band-limited to $[200,2000]$\,Hz.

Because the proposed meta-learning approach adopts few-shot training based on a handful of samples per task, the composition of the training set can influence generalization. We therefore quantify the distribution of paths with a scalar diversity score based on the average pairwise log-spectral distance (LSD). Concretely, LSD is computed on a log-spaced grid \cite{jayant1984digital} $f_m\!\in[100,2000]$\,Hz ($M{=}64$):
\vspace*{-0.1cm}
\begin{equation}
\label{eq:LSD}
\mathrm{LSD}(i,j)=\tfrac{1}{\sqrt{M}}\big\|\boldsymbol{\ell}_i-\boldsymbol{\ell}_j\big\|_2 ,
\end{equation}
\par\vspace*{-0.1cm}\noindent
\text{where}\;
$\boldsymbol{\ell}_i=\big[\,20\log_{10}|H_i(f_1)|,\ldots,20\log_{10}|H_i(f_M)|\,\big]^{\!\top}$ 
is the log-magnitude vector of path $i$. 

\begin{figure*}[!t]
  \centering
  \includegraphics[width=\linewidth]{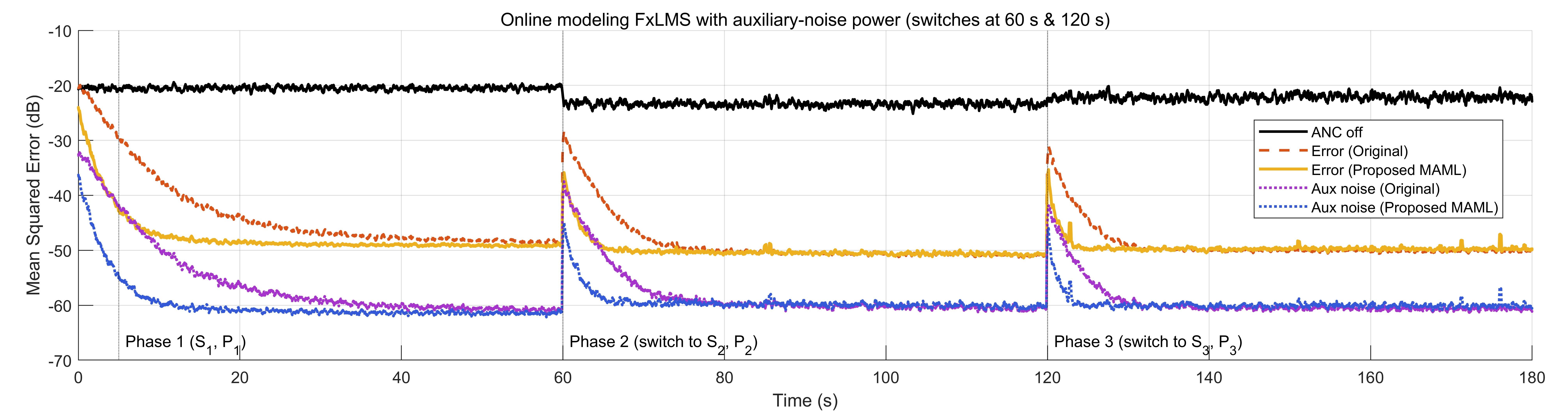}\vspace*{-0.7cm}
  \caption{Online modeling FxLMS with auxiliary-noise power (paths switch at $t{=}60$\,s \& $t{=}120$\,s).}
  \label{fig:mse-switch}
\end{figure*}

\begin{figure}[t]
  \centering
  \includegraphics[width=0.9\linewidth]{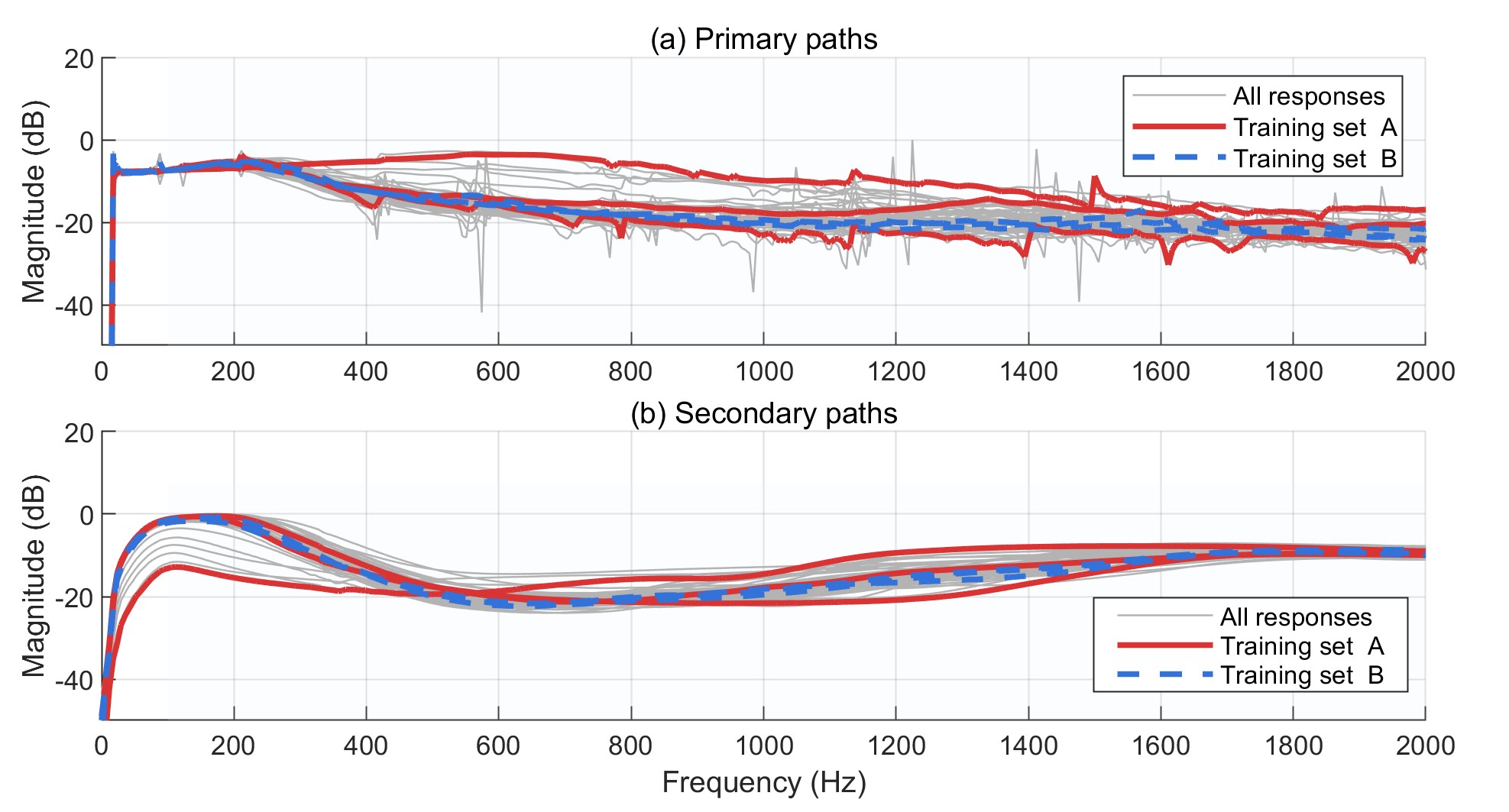}\vspace*{-0.5cm}
  \caption{Magnitude responses of the 46 measured paths from the PANDAR database.
  (a) Primary paths; (b) Secondary paths.}
  \label{fig:path-mag}
\end{figure}


To quantify how spread each subset is, we use the log-spectral distance (LSD) in \eqref{eq:LSD} and define a dispersion
\vspace*{-0.35cm}

\begin{equation}
\label{eq:disperson}
\mathcal{D}(G)
= \frac{2}{|G|(|G|-1)} \sum_{i<j,\, i,j\in G} \mathrm{LSD}(i,j).
\end{equation}
\par\vspace*{-0.2cm}\noindent
\text{where}\;
$G\subseteq\{1,\ldots,N\}$ is an index set of paths.
Figure~\ref{fig:path-mag} shows magnitude responses for all 46 pairs: thin gray curves denote all paths, while highlights mark the two training sets—A (diverse) and B (compact), each consisting of three primary–secondary pairs.
The within-set dispersion $\mathcal{D}$ is $6.17/6.75$\,dB (primary/secondary) for A and \ $0.79/0.91$\,dB for B.
\vspace*{-0.2cm}
\subsection{Evaluation of online modeling FxLMS under path change}
\vspace*{-0.2cm}
In this section, we evaluate the proposed MAML co-initialization approach with the OSPM--FxLMS in \autoref{subsec:zhang}, and testbed with 2 sudden switches of both the primary and secondary paths.
Paths are taken from the dataset mentioned in \autoref{pathdataset}, which has 46 conditions. Each condition specifies a subject and wearing fit and provides a measured primary–secondary path pair. Phase~1 runs on $(P_1,S_1)$ from \emph{person 22, slightly loose fit}; at $t{=}60$\,s the path switches to $(P_2,S_2)$ from \emph{person 13, normal fit}; at $t{=}120$\,s it switches to $(P_3,S_3)$ from \emph{person 5, normal fit}.\\
Two variants of the cross-updated scheme in \autoref{subsec:zhang} are compared:
(i) Original OSPM baseline (zero start and carry-over of $\mathbf{w}$ and $\hat{\mathbf{s}}$ across each switch), and
(ii) Proposed MAML co-init (reset at each switch).
All other settings are identical, including auxiliary noise scheduling, band-limited input noise $[200,2000]$,Hz, and random seeds.\\
\autoref{fig:mse-switch} plots the sliding MSE (dB) and the injected auxiliary-noise power. In each phase, the MAML initializer shows faster convergence and a lower early error. After both switches ($60$ and $120$ s), it reconverges more quickly and uses less auxiliary-noise energy than the baseline.

\vspace*{-0.3cm}
\subsection{Effect of training-set diversity on initialization performance}
\vspace*{-0.2cm}
To study how training-set diversity affects the MAML co-initialization , 4 training sets (A–D) was selcted, each has 3 primary-secondary path pairs with different dispersions. For each set, the dispersion $\mathcal{D}$ was computed separately on primary and secondary paths, and the residual error $5$\,s convergence was averaged over 5 unseen path pairs from dataset. The 4 sets and their dispersion are summarized in \autoref{tab:diversity-summary}.

\begin{table}[t]
\centering
\caption{Effect of training-set diversity on MAML initialization (mean noise reduction over 5 unseen conditions).}
\label{tab:diversity-summary}
\begin{tabular}{lccc}
\toprule
Training set & $\mathcal{D}_{\text{pri}}$ (dB) & $\mathcal{D}_{\text{sec}}$ (dB) & Reduction (dB) \\
\midrule
A & 6.17 & 6.75 & 26.1 \\
B & 0.91 & 0.79 & 23.2 \\
C & 5.37 & 0.67 & 23.7 \\
D & 1.55 & 4.50 & 25.8 \\
\bottomrule
\end{tabular}
\end{table}

As shown in \autoref{tab:diversity-summary}, more path diversity yields a better initializer. 
Among the two types of diversity, the secondary-path diversity matters most. 
Sets with large $\mathcal{D}_{\text{sec}}$ consistently perform better, whereas increasing only the primary-path diversity $\mathcal{D}_{\text{pri}}$ brings much smaller gains (set~C has high $\mathcal{D}_{\text{pri}}$ but low $\mathcal{D}_{\text{sec}}$ and improves little). 
This aligns with OSPM-FxLMS, where the filtered-x regressor and online modeling both depend on the secondary path. 
With a fixed data budget, prioritize collecting varied secondary paths when building few-shot training sets.

\vspace*{-0.3cm}
\section{Conclusion}
\vspace*{-0.2cm}
We propose a meta-learned co-initialization for feedforward FxLMS-based ANC, which enhances performance under changing acoustic environments by pre-training on a small set of acoustic paths. The control filter and the secondary-path model were jointly initialized through pre-training on short segments, and deployment was carried out by setting the learned initial values while the adaptive loop was left unchanged. Effectiveness under path changes was assessed on an OSPM based FxLMS testbed with measured in-ear headphone conditions. Relative to a baseline without re-initialization and with identical updates, lower early-stage residual error, shorter time to target, faster recovery after path switches, and reduced probe-noise energy were achieved. A path-distribution study indicated that the benefits of pre-training increased with training-path diversity, with secondary-path diversity exerting the strongest effect. The proposed approach complements classical FxLMS-based ANC by improving initialization, without modifying the runtime adaptive update rules.

\vspace*{-0.3cm}
\section{Acknowledgement}
\vspace*{-0.2cm}
This work was supported by the Ministry of Education, Singapore, through Academic Research Fund Tier 2 under Grant MOE-T2EP50122-0018.

\vfill\pagebreak




\apptocmd{\thebibliography}{%
  \normalsize
  \setlength{\itemsep}{1pt}
  \setlength{\parskip}{0pt}
  \setlength{\parsep}{0pt}
}{}{}

\bibliographystyle{IEEEbib}
\bibliography{strings,refs}

@article{nelson1994active,
  title={Active control of acoustic fields and the reproduction of sound},
  author={Nelson, Philip A},
  journal={Journal of Sound and Vibration},
  volume={177},
  number={4},
  pages={447--477},
  year={1994},
  publisher={Elsevier}
}

@misc{sm1996active,
  title={Active noise control system Algorithms and DSP implmentations},
  author={Kuo, Sen M and Morgan, Dennis R},
  year={1996},
  publisher={John Wiley \& Sons, INC}
}

@book{elliott2000signal,
  title={Signal processing for active control},
  author={Elliott, Stephen},
  year={2000},
  publisher={Elsevier}
}

@article{yang2023active,
  title={Active control of sound transmission through a floor-level slit},
  author={Yang, Ziyi and Wang, Shuping and Tao, Jiancheng and Qiu, Xiaojun},
  journal={The Journal of the Acoustical Society of America},
  volume={154},
  number={5},
  pages={2746--2756},
  year={2023},
  publisher={AIP Publishing}
}

@article{lam2020active,
  title={Active control of broadband sound through the open aperture of a full-sized domestic window},
  author={Lam, Bhan and Shi, Dongyuan and Gan, Woon-Seng and Elliott, Stephen J and Nishimura, Masaharu},
  journal={Scientific reports},
  volume={10},
  number={1},
  pages={10021},
  year={2020},
  publisher={Nature Publishing Group UK London}
}

@article{zhang2003robust,
  title={A robust online secondary path modeling method with auxiliary noise power scheduling strategy and norm constraint manipulation},
  author={Zhang, Ming and Lan, Hui and Ser, Wee},
  journal={IEEE Transactions on Speech and Audio Processing},
  volume={11},
  number={1},
  pages={45--53},
  year={2003},
  publisher={IEEE}
}

@article{zhang2005comparison,
  title={On comparison of online secondary path modeling methods with auxiliary noise},
  author={Zhang, Ming and Lan, Hui and Ser, Wee},
  journal={IEEE transactions on speech and audio processing},
  volume={13},
  number={4},
  pages={618--628},
  year={2005},
  publisher={IEEE}
}

@inproceedings{finn2017model,
  title={Model-agnostic meta-learning for fast adaptation of deep networks},
  author={Finn, Chelsea and Abbeel, Pieter and Levine, Sergey},
  booktitle={International conference on machine learning},
  pages={1126--1135},
  year={2017},
  organization={PMLR}
}

@article{hospedales2021meta,
  title={Meta-learning in neural networks: A survey},
  author={Hospedales, Timothy and Antoniou, Antreas and Micaelli, Paul and Storkey, Amos},
  journal={IEEE transactions on pattern analysis and machine intelligence},
  volume={44},
  number={9},
  pages={5149--5169},
  year={2021},
  publisher={IEEE}
}

@article{shi2021fast,
  title={Fast adaptive active noise control based on modified model-agnostic meta-learning algorithm},
  author={Shi, Dongyuan and Gan, Woon-Seng and Lam, Bhan and Ooi, Kenneth},
  journal={IEEE Signal Processing Letters},
  volume={28},
  pages={593--597},
  year={2021},
  publisher={IEEE}
}

@article{hu2019online,
  title={Online multi-channel secondary path modeling in active noise control without auxiliary noise},
  author={Hu, Meiling and Xue, Jinpei and Lu, Jing},
  journal={The Journal of the Acoustical Society of America},
  volume={146},
  number={4},
  pages={2590--2595},
  year={2019},
  publisher={AIP Publishing}
}

@book{liebich2019acoustic,
  title={Acoustic path database for ANC in-ear headphone development},
  author={Liebich, Stefan and Fabry, Johannes and Jax, Peter and Vary, Peter},
  year={2019},
  publisher={Universit{\"a}tsbibliothek der RWTH Aachen}
}

@inproceedings{wang2025transferable,
  title={Transferable Selective Virtual Sensing Active Noise Control Technique Based on Metric Learning},
  author={Wang, Boxiang and Shi, Dongyuan and Luo, Zhengding and Shen, Xiaoyi and Ji, Junwei and Gan, Woon-Seng},
  booktitle={ICASSP 2025-2025 IEEE International Conference on Acoustics, Speech and Signal Processing (ICASSP)},
  pages={1--5},
  year={2025},
  organization={IEEE}
}

@article{luo2023delayless,
  title={Delayless generative fixed-filter active noise control based on deep learning and bayesian filter},
  author={Luo, Zhengding and Shi, Dongyuan and Gan, Woon-Seng and Huang, Qirui},
  journal={IEEE/ACM Transactions on Audio, Speech, and Language Processing},
  volume={32},
  pages={1048--1060},
  year={2023},
  publisher={IEEE}
}

@article{kuo2006active,
  title={Active noise control system for headphone applications},
  author={Kuo, Sen M and Mitra, Sohini and Gan, Woon-Seng},
  journal={IEEE Transactions on Control Systems Technology},
  volume={14},
  number={2},
  pages={331--335},
  year={2006},
  publisher={IEEE}
}

@book{jayant1984digital,
  title={Digital coding of waveforms: principles and applications to speech and video},
  author={Jayant, Nuggehally S and Noll, Peter},
  volume={2},
  year={1984},
  publisher={Prentice-Hall Englewood Cliffs, NJ}
}

@article{kajikawa2012recent,
  title   = {Recent advances on active noise control: open issues and innovative applications},
  author  = {Kajikawa, Yoshinobu and Gan, Woon-Seng and Kuo, Sen M.},
  journal = {APSIPA Transactions on Signal and Information Processing},
  volume  = {1},
  pages   = {e3},
  year    = {2012},
  doi     = {10.1017/ATSIP.2012.4}
}

@article{sun2020realistic,
  title   = {A Realistic Multiple Circular Array System for Active Noise Control over 3D Space},
  author  = {Sun, Huiyuan and Abhayapala, Thushara D. and Samarasinghe, Prasanga N.},
  journal = {IEEE/ACM Transactions on Audio, Speech, and Language Processing},
  volume  = {28},
  pages   = {3041--3052},
  year    = {2020},
  doi     = {10.1109/TASLP.2020.3038551}
}

@article{zhang2018wave,
  title = {Active Noise Control Over Space: A Wave Domain Approach},
  author = {Zhang, Jihui and Abhayapala, Thushara D. and Zhang, Wen and Samarasinghe, Prasanga N. and Jiang, Shouda},
  journal = {IEEE/ACM Transactions on Audio, Speech, and Language Processing},
  volume = {26},
  number = {4},
  pages = {774--786},
  year = {2018},
  doi = {10.1109/TASLP.2018.2795756}
}

@article{zhang2021deep,
  title={Deep ANC: A deep learning approach to active noise control},
  author={Zhang, Hao and Wang, DeLiang},
  journal={Neural Networks},
  volume={141},
  pages={1--10},
  year={2021},
  publisher={Elsevier}
}

@inproceedings{schumacher2011active,
  title={Active noise control in headsets: A new approach for broadband feedback ANC},
  author={Schumacher, Thomas and Kr{\"u}ger, Hauke and Jeub, Marco and Vary, Peter and Beaugeant, Christophe},
  booktitle={2011 IEEE International conference on acoustics, speech and signal processing (ICASSP)},
  pages={417--420},
  year={2011},
  organization={IEEE}
}

@article{lopes2017close,
  title={A close to optimal adaptive filter for sudden system changes},
  author={Lopes, Paulo AC and Gerald, Jos{\'e} AB},
  journal={IEEE Signal Processing Letters},
  volume={24},
  number={11},
  pages={1734--1738},
  year={2017},
  publisher={IEEE}
}

@article{yang2017frequency,
  title={Frequency-domain adaptive Kalman filter with fast recovery of abrupt echo-path changes},
  author={Yang, Feiran and Enzner, Gerald and Yang, Jun},
  journal={IEEE Signal Processing Letters},
  volume={24},
  number={12},
  pages={1778--1782},
  year={2017},
  publisher={IEEE}
}

@inproceedings{monteiro2017accelerating,
  title={Accelerating the convergence of adaptive filters for active noise control using particle swarm optimization},
  author={Monteiro, Rodrigo P and Lima, Gabriel A and Oliveira, Jos{\'e} PG and Cunha, Daniel SC and Bastos-Filho, Carmelo JA},
  booktitle={2017 IEEE Latin American Conference on Computational Intelligence (LA-CCI)},
  pages={1--6},
  year={2017},
  organization={IEEE}
}

\end{document}